# Effect of spoke design and material nonlinearity on non-pneumatic tire stiffness and durability performance

Priyankkumar Dhrangdhariya[1], Soumyadipta Maiti[1*], Beena Rai[1]

[1] TCS Research, Tata Research Development and Design Centre, 54-B, Hadapsar Industrial Estate, Pune-411013, India

* Corresponding author email: soumya.maiti@tcs.com

## Abstract

Non-pneumatic tire has been widely used due to their advantages of no run-flat, no need of air maintenance, low rolling resistance, and improvement of passenger's comfort due to its better shock absorption. It has variety of application in the military vehicle, earthmovers, lunar rover, stair climbing vehicles etc. Recently UPTIS (Unique Puncture-Proof Tire System) non pneumatic tire has been introduced for passenger vehicles. In this study three different design configuration Tweel, Honeycomb and newly developed UPTIS have been compared. Effect of Polyurethane (PU) material nonlinearity have also been introduced by applying 5 different nonlinear PU material property in the spokes. The combined analysis of the PU material nonlinearity and spoke design configuration on the overall tire stiffness and spoke damage prediction is analysed using 3-Dimensional FEM simulations performed in ANSYS 16.0. It has been observed that Mooney Rivlin 5-parameter model is best to capture all 5 studied PU material's the nonlinearity. Effect of material nonlinearity on various spoke designs have been studied. The best combination of spoke design and the use of nonlinear material have been suggested in terms of riding comfort, tire stiffness and durability performance.

## 1 Introduction

For last few years, the automotive industries are looking into the development of non-pneumatic tyres (NPTs) to attain the potential advantages over pneumatic tyres in terms of no run-flat, no need of air maintenance, low rolling resistance and improvement of passenger's comfort due to better shock absorption [1–3]. Several tire engineers have attempted to develop non-pneumatic tires by filling an elastomer into the space for air or by building polygon typed spokes to replace the air of a pneumatic tire [3–6].

A typical NPT usually consists of a hub, flexible spokes, a shear ring and a tread. The axle of the vehicle is attached to the rigid hub. Aluminium alloy is used as a material for Hub. The main task of the tread is to provide proper cooperation of its blocks with the ground and providing protection of reinforcement layers against damage resulting from the impact of road's obstacles. A rubber compound is used as a material for a tread layer. The shear ring acts as a supporting structure determines the properties of the non-pneumatic tire (e.g. directional stiffness, rolling resistance), which consists of composite structure composed of a shear band with two circumferential reinforcements (i.e., inner and outer rings)[3,7,8]. Flexible spokes have remarkable impact on the overall performance of NPT, which is made from Polyurethane[9]. The NPT stiffness properties and contact pressure during loading is affected by several geometric and material parameters, including ring thickness, reinforcement type, and spoke geometry [10]. Various shapes of flexible spokes have been introduced by Michelin, Bridgestone, and Polaris for the desired application like Military Vehicle, Earthmovers, Lunar Rover, Stair climbing vehicles, Passenger vehicles etc. In 2005, Michelin developed the Non pneumatic tire named as Tweel which is classified by Time magazine as "one of the most amazing inventions" in that year[11]. Resilient Technology introduced NPT having Honeycomb spokes in 2012, later tested and used by U.S army in the military vehicle for its excellent 'bulletproof' capability[12,13]. In 2013, Bridgestone revealed its second-generation air free concept non-pneumatic tire featuring improved load-bearing capabilities, environmental design and driving performance. Recently Michelin presented UPTIS (Unique Puncture-Proof Tire System) at the Movin'On Summit in 2017, which is an airless mobility solution for passenger vehicles with the goal of introducing the model as early as 2024[14].

Many literatures are available in design of spokes and its effect on the cellular spoke or variation in honeycomb spoke design [7,15–17]. Kucewicz et al presented the modelling methodology of non-pneumatic tire having different spoke structure and conducted radial deflection analysis using Finite element modelling (FEM) techniques[17]. Ingrole et al compared different Honeycomb spoke designs and suggested Auxetic structure having good compressive strength compared to other designs [15]. The team has found good agreement of spoke deformation shape received from FEM with experiments. Bezgam et al analysed the effect of cell angles of honeycomb spokes with the same thickness and same load carrying capacity on the deformation, stress distribution and contact pressure distribution of NPT[18]. Ju et al investigated the validity of NPT with high fatigue resistance by achieving low local stress under vertical stiffness loading using FEM analysis. The elastic limits of several hexagon honeycomb structure having different cell angle of spokes were analysed and compared to suggest a better geometrical designs of lower localized peak stress and fatigue properties [7]. In literature three different configurations of Michelin Tweel, Resilient technology and Bridgestone NPT were investigated by conducting 2D FEM analysis, to find out comparison of these spoke structure on the contact pressure, vertical tire stiffness and stresses of NPT. The results showed that the shape of the spokes have a great effect on tire's behaviour. [19]

Macro and local cellular properties of hexagonal honeycombs have also been extensively studied[10,20–23]. Rugsaj et al have extracted the material property of NPT by preparing the specimen using water jet cutting. They have developed 3D FEM model to capture the deformation behaviour of these NPT component with the help of hyperelastic constitutive models (i.e. Neo Hookean model, Ogden model, Mooney Rivlin model)[24]. The most commonly used material models for describing the properties and behavior of hyperelastic material in supporting structure, spokes and tread are: Ogden, Mooney-Rivlin, Neo-Hookean and Marlow [7,8,17,25].

There are many factors which are useful to understand the durability performance of NPT. The strain energy density (SED) has been used as a predictor of fatigue life in elastomers since the development of the fracture mechanics for this class of material [26]. The SED has been found to be proportional to crack density and size under simple load conditions, so it can be used as a measure of the fatigue life of the material [26]. Abraham et. al, has also pointed out that SED is a better predictor than maximum strain or stress based predictors [27] . Other than SED, Maximum Principal strain/stretch and octahedral shear strain can also be used as crack initiation predictor for the hyper-elastic material. [28]. Stress, apparently on the other hand has rarely been used as a fatigue life parameter [29].

It has been found that several researchers are taking the elastic modulus of the PU as constant (E ~32 MPa) which was applied for shear beam and spokes [7,19]. But in reality, PU material properties can become highly non-linear depending on the material processing routes. In this study, we have focused to emphasize on the effect of non-linearity for PU stress-strain behaviour which have : (1) Same Elastic rubber modulus around standard 300% strain but different non-linear stress-strain behaviour (2) Different elastic modulus at 300% strain but have similar nonlinearity of stress-strain; on the Tire stiffness and Durability analysis. In this study the comparative analysis of Michelin UPTIS (which is recent development of Michelin for Passenger Vehicle) with Honeycomb and Tweel spokes in terms of Tire stiffness and durability performance (with the help of SED, Principal strain and Octaherdal shear strain) has been done, which is rarely found in the literature. Based on the FEM simulation results, we compared the performance of three different spokes and 5 different nonlinear PU property and suggested the optimum combination of PU property and spoke structure in terms of better durability, stiffness and riding comfort.

## 2 Modeling & Simulation Details

### 2.1 CAD Model Details:

As mentioned in the introduction section, NPT consists of a Hub, Shear band with having inner and outer reinforcement rings, Flexible spokes and tread. The detailed description of NPT is mentioned in the Figure 1.

Since none of the companies provide information about materials used and dimensions, all executed numerical models were developed based on articles, photos and press reports which are available on the literature [17]. In this study we selected the dimensions of the non-pneumatic tyre's components based on that are available in the literature [7,19].

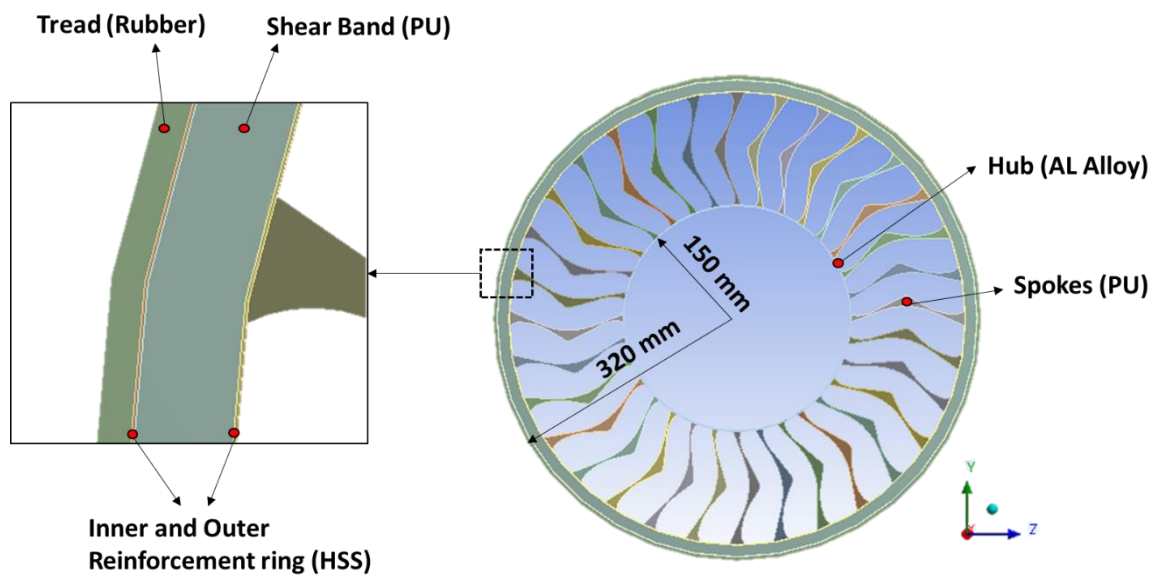

Figure 1. Detailed description of Non Pneumatic Tire

We developed CAD models which depicts three different NPT models i.e (a) Michelin UPTIS (b) Michelin X-Tweel and (c) Resilient's Honeycomb using Free-CAD 0.18 as shown in the Figure 2. Then, the CAD drawings are it converted to '.step' file format and imported to ANSYS 16.0 for further analysis. Here dimensions of all the components are similar except spoke structure. Spoke's thickness and numbers have been selected such that the volume difference of all three spokes are less than 5% and the volume of all spokes is almost equivalent and comparable. Diameter of Hub and outside dimensions of NPT are taken as shown in the Figure 1. The in-plane thickness of NPT is 215 mm. Thickness of the hub, inner and outer reinforcement rings, shear band and treads are 1 mm, 0.5 mm, 15 mm and 5 mm respectively.

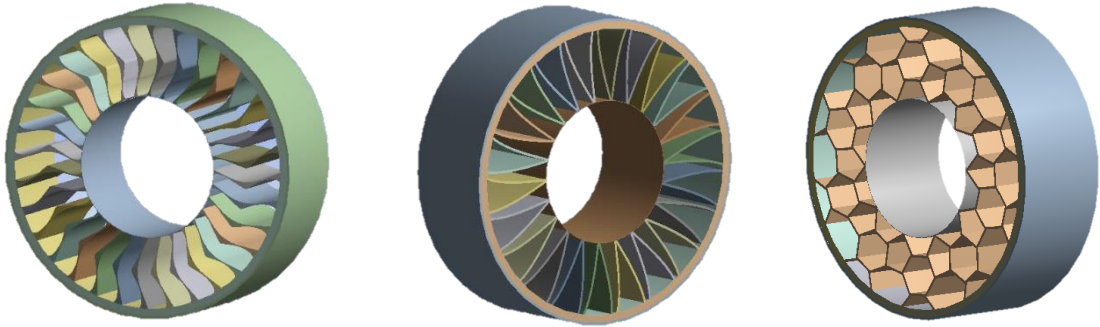

Figure 2. CAD Models of Non-Pneumatic tire having different spoke design

(a) UPTIS        (b) Tweel        (c) Honeycomb

## 2.2  Material property and Hyperelastic curve fitting

Material property for the hub, inner and outer reinforcement ring, and tread have been selected as shown in Table 2.1. [7].

**Table 2.1 Material property of NPT Components**

| *Components* | *Materials* | *Density (kg/m$^3$)* | *E* | *μ* |
|---|---|---|---|---|
| Hub | 7075 T6 Al Alloy | 2800 | 72 Gpa | 0.33 |
| Inner and Outer Ring | ANSI 4340 HSS | 7800 | 210 Gpa | 0.29 |
| Tread | Rubber | 1043 | 11.9 MPa | 0.49 |

As a general practice Polyurethane (PU) is used as a suitable material for the spokes and the shear beam. In this study we have extracted 5 different PU stress-strain data from Cristina et al and Milena et al. The behaviour of these 5 various PU stress-strain property and their hyper elastic curve fitting are as follows [30,31].

### 2.2.1  *Material property of spokes and shear beam*
**Case 1)** Having similar 300% modulus and Different nonlinearity:

The engineering stress-strain data for 4,4'-diphenyl methane diisocyanate (MDI) and 4,4'-dibenzyl diisocyanate (DBDI) are available in the Cristina et al as shown in the Figure 3a. The difference in the MDI and DBDI is degree of crystallinity, MDI is non-crystallizing and DBDI is crystallizing (14% degree of crystallinity). Here MDI and DBDI, both have equal ~300 % rubber modulus but the curvature of the engineering stress-strain curve is different [31].

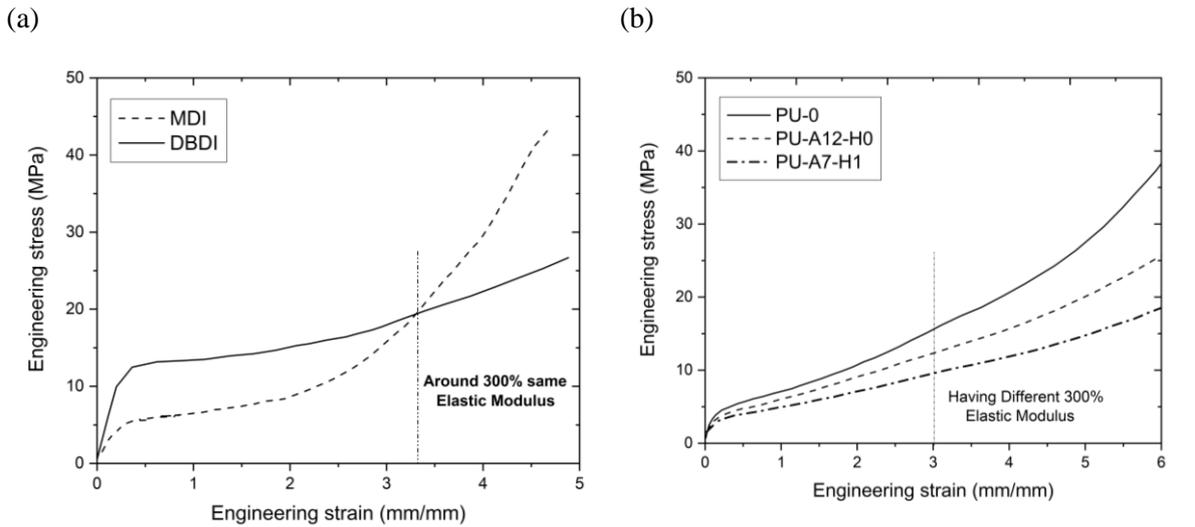
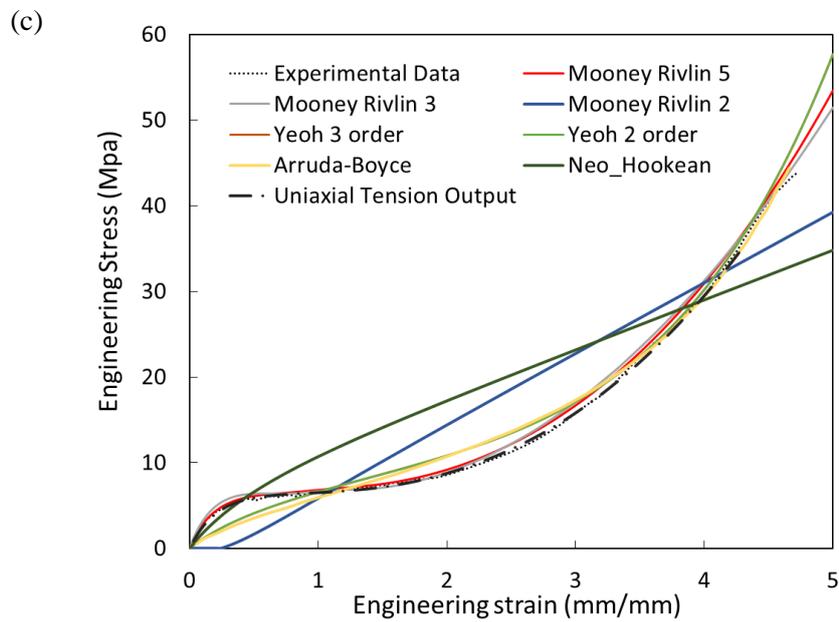
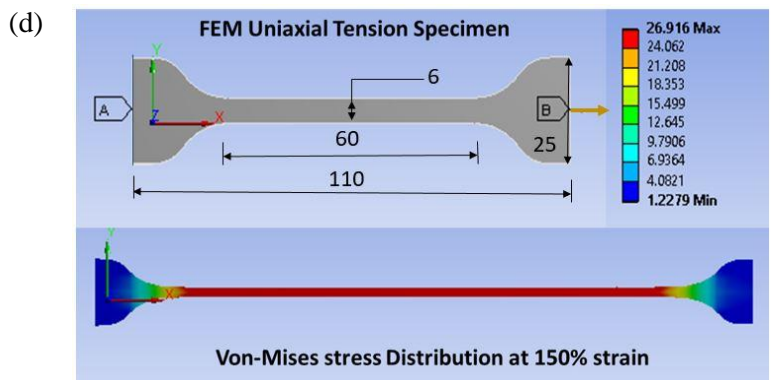

(All Dimensions are in mm)

Figure 3. (a,b) 5 different nonlinear PU material property stress strain Curve (c) MDI Curve fitting using Hyperelastic models (d) FEM simulation of Uniaxial tensile specimen

**Case 2)** Having Same Linearity and Different 300% Modulus:

The engineering stress-strain data have been plotted for three different PU (PU-0, PU-A12-H0 and PU-A7-H1) having almost similar curvature (or linearity) but different 300% elastic modulus as shown in Figure 3b. These have different size of filler particle, weight percentage of hard segment content and degree of crystallinity as shown in the Table 2.2 [30].

**Table 2.2 Characteristics of PU Nanocomposites**

| Code | Silica Nanofiller | Filler size (nm) | Hard segment content (%) | Degree of crystallinity (%) |
|---|---|---|---|---|
| *PU-0* | None | - | 17.8 | 7.1 |
| *PU-A12-H0* | Aerosil 380 (A12) | 12 | 17.4 | 9.3 |
| *PU-A7-HI* | Aerosil R974 (A7) | 07 | 17.7 | 8.8 |

### 2.2.2 Hyperelastic model curve fitting

To capture these non-linear behaviour of 5 different PU property, we need to find the hyperelastic constants of the hyperelastic models as mentioned below. [32]

1. *Mooney-Rivlin Hyperelasticity:*

The strain energy function ($\psi$) of a Mooney-Rivlin hyperelastic model can be expressed as mentioned in the equation 1.

$$\psi = C_{10}(\bar{I}_1 - 3) + C_{01}(\bar{I}_2 - 3) + C_{11}(\bar{I}_1 - 3)(\bar{I}_2 - 3) + C_{20}(\bar{I}_1 - 3)^2 + C_{02}(\bar{I}_2 - 3)^2 + \frac{1}{d}(J - 1)^2$$
(1)

Where $\bar{I}_{1,2}$ is the deviatoric first principal invariant, J is the Jacobian and the required input parameters are $C_{10}$, $C_{01}$, $C_{11}$, $C_{20}$, $C_{02}$ (material constants) and d (material incompressibility parameter). The initial shear modulus is defined as ($\mu = 2(C_{10} + C_{01})$) and the initial bulk modulus is defined as (K = 2/d).

The equation 1 have 5 material constants and indicates Mooney-Rivlin 5 parameter model, while Mooney-Rivlin 2 parameter model doesn't have 3rd, 4th and 5th term, and Mooney-Rivlin 3 Parameter model doesn't count 4th and 5th term.

2. *Yeoh Hyperelasticity:*

   The Yeoh hyperelastic strain energy function ($\psi$) is similar to the Mooney-Rivlin models described above except that it is only based on the first deviatoric strain invariant ($\bar{I}_1$). It has the general form as mentioned in the equation 2.

$$\psi = \sum_{i=1}^{N} C_{i0} (\bar{I}_1 - 3)^i + \sum_{k=1}^{N} \frac{1}{d_k} (J-1)^{2k} \qquad (2)$$

Where N= order of Yeoh Hyperelastic models (i.e 1, 2, 3).

In Yeoh 3rd order hyperelastic model, required input parameters are three material constants ($C_{10}$, $C_{20}$, $C_{30}$) and three incompressibility parameters ($d_1$, $d_2$, $d_3$). While, In Yeoh 2nd order hyperelastic model, required input parameters are two material constants ($C_{10}$, $C_{20}$) and two incompressibility parameters ($d_1$, $d_2$). The initial shear modulus is defined as $\mu = 2 C_{10}$ and the initial bulk modulus is defined as $K = 2/d_1$.

3. *Neo Hookean Hyperelasticiy:*

The strain energy function ($\psi$) for the Neo-Hookean hyperelastic model is shown in the equation 3. Where the required input parameters are initial shear modulus of the material ($\mu$) and incompressibility parameter (d). The initial bulk modulus (K) is defined as $K = 2/d$.

$$\psi = \frac{\mu}{2} (\bar{I}_1 - 3) + \frac{1}{d} (J-1)^2 \qquad (3)$$

4. *Arruda-Boyce Hyperelasticiy:*

The strain energy function ($\psi$) of Arruda-Boyce hyperelastic model is shown in equation 4.

$$\psi = \mu \left[ \frac{1}{2} (\bar{I}_1 - 3) + \frac{1}{20 \lambda_L^2} (\bar{I}_1^2 - 9) + \frac{11}{1050 \lambda_L^4} (\bar{I}_1^3 - 27) + \frac{19}{7050 \lambda_L^6} (\bar{I}_1^4 - 81) + \frac{519}{673750 \lambda_L^8} (\bar{I}_1^5 - 243) \right] + \frac{1}{d} \left( \frac{J^2-1}{2} - \ln J \right) \qquad (4)$$

Where the required input parameters are initial shear modulus of the material ($\mu$), limiting network stretch ($\lambda_L$), and incompressibility parameter (d). The initial bulk modulus (K) is defined same as mentioned in the Neo-Hookean hyperelasticity.

To check the best hyperelastic model curve fitting, we selected above hyper elastic models and applied to MDI type PU stress-strain data. The required input parameters have been calculated using ANSYS 16.0 curve fitting tool as shown in the table. As the material is incompressible, the incompressibility parameter selected as tends to zero value. The curve fitting of these hyperelastic models to the MDI experimental data has been plotted in the figure 3c.

**Table 2.3 Value of required input parameters calculated from ANSYS 16.0 curve fitting tool for MDI experimental data**

| Model/ Constants | C10 | C01 | C20 | C11 | C02 | C30 | $\mu$ | $\lambda_L$ |
|---|---|---|---|---|---|---|---|---|
| *Mooney Rivlin 5* | -3.883 | 9.3616 | 0.0626 | 0.0578 | 0.5533 | - | - | |
| *Mooney Rivlin 3* | -4.9677 | 11.306 | - | 4.9994 | - | - | - | - |
| *Mooney Rivlin 2* | 4.1205 | -5.1963 | - | - | - | - | - | - |
| *Yeoh 3rd Order* | 1.8131 | - | 0.00107 | - | - | -0.011 | - | - |
| *Yeoh 2nd Order* | 1.21 | - | 0.04192 | - | - | - | - | - |
| *Arruda-Boyce* | - | - | - | - | - | - | 3.0412 | 3.5724 |
| *Neo Hookean* | - | - | - | - | - | - | 5.787 | - |

To identify the quality of the curve fitting, Sum squared error (or Residual) value and $R^2$ value have been calculated as mentioned in the **Error! Reference source not found.**. $R^2$ value is almost equal and tends to 0.99 for Mooney Rivlin 5 parameter, 3 parameter, Yeoh-3rd order, Yeoh 2nd order, and Arruda-Boyce model. But If we compare the Sum square error or Residual value then we can identify that Mooney-Rivlin 5 parameter model has lowest value of residual and hence we have selected the Mooney-Rivlin 5 parameter model for the rest of the Polyurethane (i.e DBDI, PU-0, PU-A12-H0, and PU-A7-H1) curve fitting. The Material constants are mentioned in the Table 2.4. To verify the curve fitting quality of Mooney Rivlin 5 parameter model, FEM uniaxial tensile specimen (dog-bone shape) has been designed as shown in fig 3d. Thickness of the specimen has been taken as 2 mm. All the dimension are shown in the figure. SOLID 186 (3D 20 node Hexahedral and tetrahedral element) are used to mesh the model. Left hand side of the specimen (mention as A) has applied fixed support and around 150% strain is applied to the right side of the specimen (mentioned as B) as shown in fig 3d. von Mises stress distribution at 150% strain is shown in the figure. It can be observed from the von Mises stress contour that the stress distribution is very uniform throughout the gauge length of the tensile test specimen. The FEM output of uniaxial tension has also been presented in the fig 3c, and it is giving exact match with input MDI engineering stress-strain curve. Hence it can be analysed that the Mooney-Rivlin 5 parameter hyperelastic model is giving best curve fit to capture PU material non-linearity.

**Table 2.4 Residual (SSE) and R2 calculation of hyperelastic curve fitting**

| Hyperelastic Models | Residual (SSE) (MPa) | $R^2$ |
|---|---|---|
| Mooney Rivlin 5 | 6.077 E+06 | 0.99979 |
| Mooney Rivlin 3 | 1.729 E+07 | 0.99941 |
| Mooney Rivlin 2 | 1.447 E+09 | 0.95098 |
| Yeoh 3$^{rd}$ Order | 2.721 E+08 | 0.99078 |
| Yeoh 2$^{nd}$ Order | 1.127 E+08 | 0.99618 |
| Arruda-Boyce | 1.421 E+08 | 0.99518 |
| Neo-Hookean | 1.972 E+09 | 0.93471 |

**Table 2.4 Mooney-Rivlin 5 Parameter model's material constants for PU**

| Material | C10 | C01 | C20 | C11 | C02 |
|---|---|---|---|---|---|
| MDI | -3.883 | 9.3616 | 0.0626 | 0.0578 | 0.5533 |
| DBDI | -16.565 | 30.572 | -0.0281 | 0.3005 | 3.7316 |
| PU-0 | -24.831 | 34.049 | 0.1064 | -0.8767 | 9.0086 |
| PU-A12-H0 | -17.014 | 24.208 | 0.0431 | -0.3908 | 5.8038 |
| PU-A7-H1 | -11.817 | 17.440 | 0.0253 | -0.2277 | 3.9123 |

## 2.3 FEM Simulation Details:

CAD models of NPT-UPTIS, Tweel and Honeycomb structures have been imported to ANSYS 16.0 Workbench for the FEM Simulation. All interfaces, the spokes, hub, inner and outer reinforcement ring, shear band, and tread are connected to each other using bonded contact. Road surface and Tread of the tire has been tied to each other with rough contact which won't allow sliding between tread and road surface.

Fixed support constraint is applied to the inner face of the hub, which restricts all translational and rotational movement. A 50 mm of displacement has been applied normal to the road surface as shown in the Figure 4. The 50 mm of road displacement is applied with 2 mm increment with total 25 gradual steps to reduce the element distortion and convergence error. SOLID 187 (3D 10 node Tetrahedral element) used to model thin reinforcement rings and SOLID 186 (3D 20 node Hexahedral and tetrahedral element) are used in the spokes, tread, shear band, hub and road surfaces. TARGE170 and CONTA174 have been used to identify bonded target and contact interfaces, (Speciality of bonded contact). Due to large deformation and material nonlinearity it was necessary to use nonlinear method to solve FEM equations. Therefore, iterative Newton-Raphson solution was used in this FEM simulation.

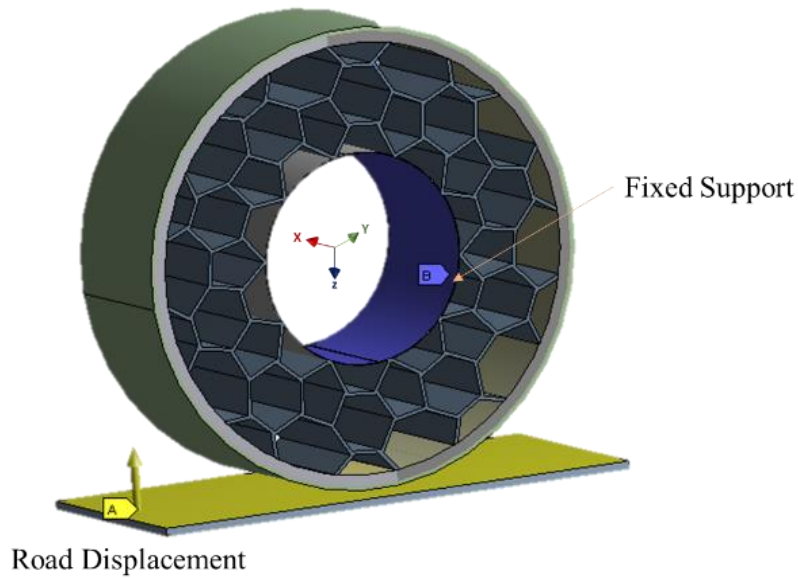

Figure 4. Boundary conditions description

## 3  Results & Discussion

### 3.1  Strain distribution for three types of spokes

Figure 4 shows von-Mises strain distribution of Honeycomb, Tweel and UPTIS spokes having MDI as a spoke's material property and deformed by 50 mm of road displacement. It is observed that the lower portion of the spokes is more strained, indicated by red box. However, the region of maximum von Mises strain depends much on the spoke design. For example, for Honeycomb and Tweel, this region is close to the Tread part, whereas it is close to the rim for UPTIS. It has also been analysed that at the same degree of road displacement, the von-Mises strain is higher in the UPTIS spoke design compared to Honeycomb and Tweel structure. This has also been the general tendency observed among different spoke designs when loaded with various other linear and non-linear PU properties.

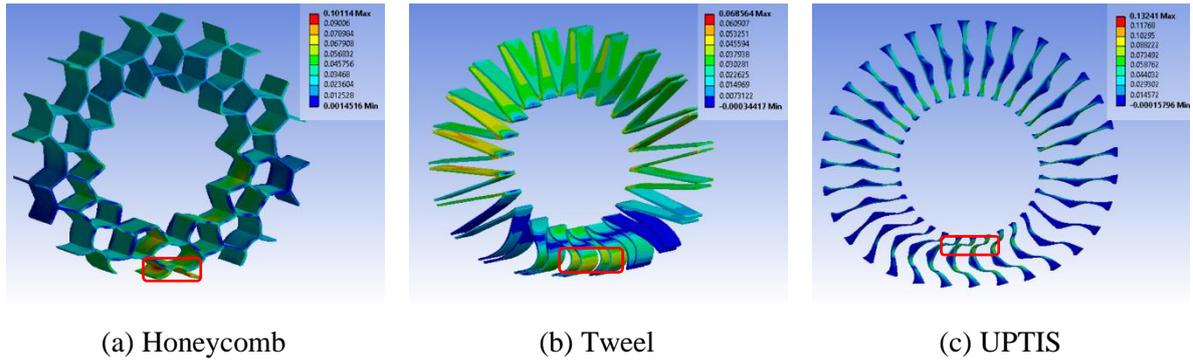

(a) Honeycomb          (b) Tweel          (c) UPTIS

Figure 4. Von-Mises strain distribution in the various spoke structure

## 3.2 Effect of PU material non-Linearity and spoke design on Tire stiffness and Damage prediction.

### *3.2.1 Tire Stiffness:*

Fig 5 shows the static load displacement analysis for (a) Linear elastic PU (Having constant elastic modulus of 32 MPa) [7,19] and (b) MDI (Highly nonlinear PU, having different elastic modulus at different percentage of strain) as discussed earlier in the Fig 3a. It is observed that both linear elastic PU and Nonlinear MDI both have quite similar amount of stiffness. Although all three spoke designs have equivalent spoke volume, UPTIS spoke design is less stiff or having more cushioning effect compared to Tweel and Honeycomb spokes. By comparing the nature of the static load displacement curve, it can be seen that Tweel and Honeycomb have almost linear or constant stiffness slope at higher and lower loading, while UPTIS has different degree of stiffness at lower and higher loading value. The percentage difference in tire stiffness for the different spoke designs goes down with load increment. For the first 10 mm of displacement, the overall tyre stiffness for UPTIS is around 41% less than the Honeycomb and Tweel counterpart. The stiffness difference goes down to 20-30% range among the spoke designs at higher displacement of 40 mm.

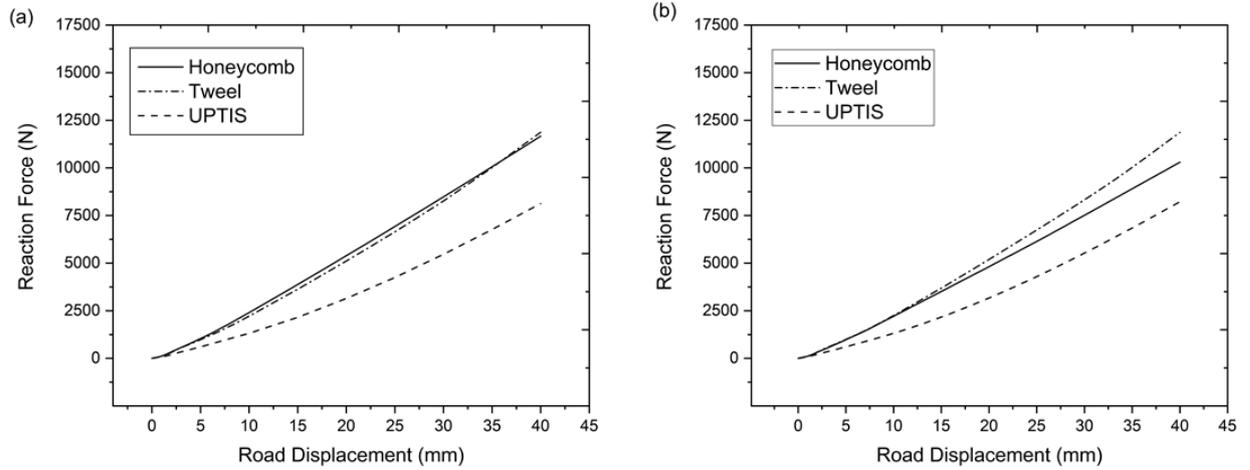

Figure 5 Static load displacement curves of various spoke designs having (a) Linear Elastic Poly Urethane (E=32MPa) (b) Non-Linear Poly Urethane (MDI) material property for spokes

Fig 6 shows the static load displacement analysis of NPT-Tweel for (a) MDI, DBDI (having around same 300% elastic modulus but different non-linearity) and (b) PU-0, PU-A12-H0, and PU-A7-H1 (Having different 300% elastic modulus but similar non-linearity) material property applied for spokes and shear beam while keeping all other material property as constant. The similar tire stiffness behaviour for these 5-material properties has also been observed for NPT-Honeycomb and NPT-UPTIS as well. For the sake of simplicity NPT-Tweel load displacement curve has been shown as the stiffness difference in the MDI and DBDI is more pronounced. From fig 6(a) it has been illustrated that the non-linearity of PU property has significant effect on the tire stiffness. In Fig 6 (b) it can be seen that, As PU-0, PU-A12-H0 and PU-A7-H1 have similar linearity but having different 300% elastic modulus, the stiffness trend of these three material properties are similar to their material stress-strain nature. From Fig. 6(a) it is evident that the tyre stiffness difference due to material non-linearity is almost 110%, whereas there is virtually no difference of the rubber modulus at a standard 300% stretch. In Fig. 6(b) the difference in tyre stiffness at under 10 mm displacement is around 17%, which is even little smaller to the 21% difference found in the rubber modulus at a standard 300% stretch shown in Fig. 3a2.

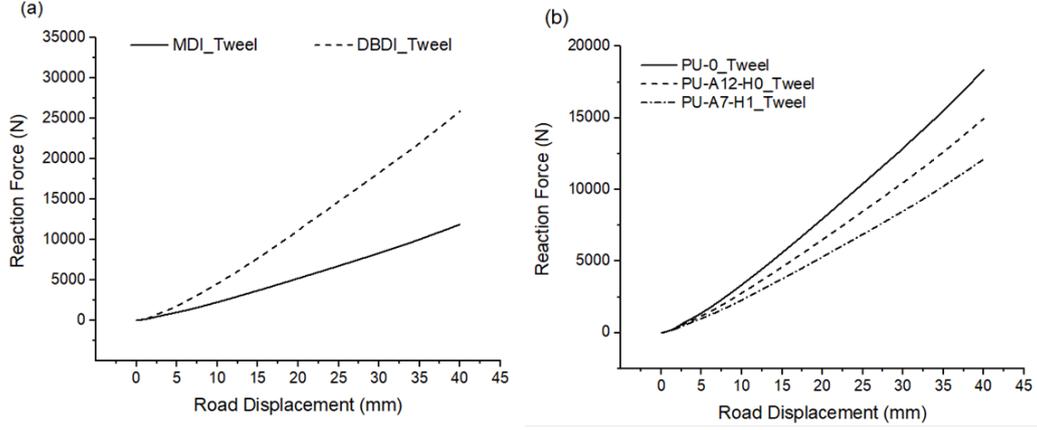

Figure 6 Static load displacement analysis for NPT having Tweel spokes and having material property of (a) MDI, DBDI and (b) PU-0, PU-A12-H0, and PU-A7-H1. Effect of tyre stiffness depending on material properties.

### *3.2.2 Damage Prediction and Durability performance:*

As discussed in section 2, SED, principal strain and octahedral shear strains are used to predict the crack initiation in the elastomers, which is helpful to analyse the damage prediction of non pneumatic tire in spokes. [26,28]

SED and octahedral shear strain can be calculated as described in the following equation 5 & 6 respectively.

$$\text{SED} = \frac{1}{2}\ [\sigma_x*\varepsilon_x + \sigma_y*\varepsilon_y + \sigma_z*\varepsilon_z + \sigma_{xy}*\varepsilon_{xy} + \sigma_{yz}*\varepsilon_{yz} + \sigma_{xz}*\varepsilon_{xz}] \tag{5}$$

$$\text{Octahedral shear strain} = \frac{2}{3}\sqrt{[(\varepsilon_1-\varepsilon_2)^2 + (\varepsilon_2-\varepsilon_3)^2 + (\varepsilon_1-\varepsilon_3)^2]} \tag{6}$$

In Fig 7(a) and 7(c) SED variation with respect to reaction force in spokes is illustrated for LE Polyurethane and MDI. Similarly, maximum principal strain variation in the spokes is illustrated in the Fig 7(b) and Fig 7(d).

Fig 7 shows that the value of strain energy density (SED) and max principal strain is lowest in the Tweel spoke design and highest in the UPTIS. There is quick rise in the SED and Max principal strain for the Honeycomb spoke after around 400 kg of static load, which is due to the highly localized elastic deformation in the honeycomb spokes due to its structural design. By comparing SED curve for the linear elastic PU property (Fig 7a) and nonlinear PU (MDI) (fig 7c), it can be seen that irrespective of the material non-linearity, there can be a large deviation of important damage parameter like SED and Tweel seems to be the better design compared to the other spoke designs. We may infer that Tweel has more crack initiation resistance than Honeycomb and UPTIS spoke structure.

Having identified Tweel as a design of choice from a damage initiation perspective, we have investigated the effect of various material properties and selection of damage parameters like SED, maximum principal strain and octahedral shear strain. From Fig 8(a), it can be illustrated that the value of SED is almost equal till 1000 kg of static load for NPT-Tweel for all the 5 material properties, and at higher load DBDI has less amount of SED value than all other material properties. Maximum principal strain and maximum octahedral shear strain variation are shown in the Fig 8(b) and fig 8(c). MDI and PU-A7-H1, both have almost similar value of SED, maximum principal strain and maximum octahedral shear strain. If we look at all the crack initiation parameter, we can observe that the DBDI has more crack initiation resistance and hence good durability performance. As SED predicts similar values of damage for all materials until a load of 1000 N, the other damage parameters like maximum strain and maximum octahedral strains appear to be useful to distinguish the material with least damage factors. Here DBDI becomes the favourable candidate material according to this study.

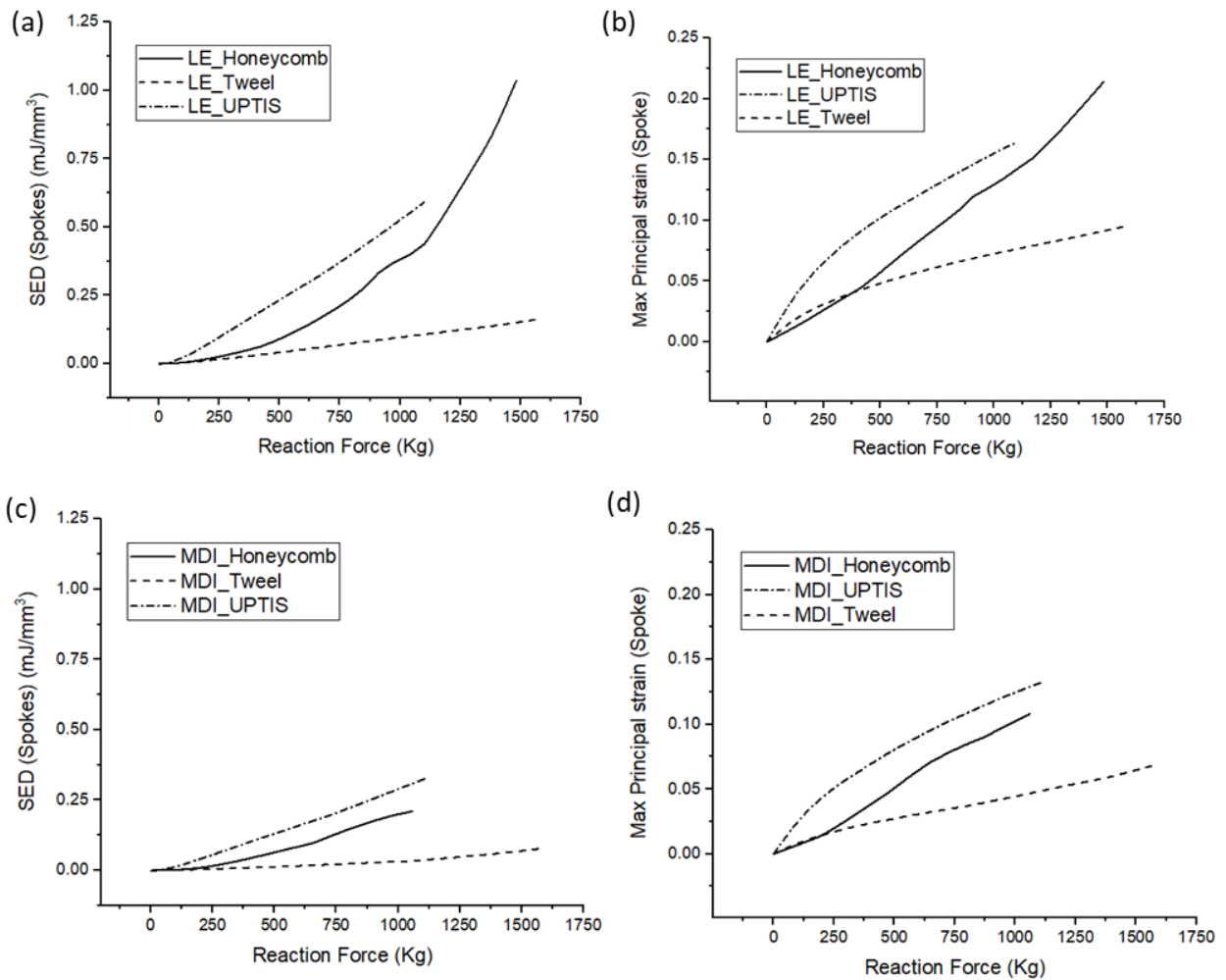

Figure 7 SED & maximum principal strain vs reaction force for (a, b) LE Polyurethane and (c, d) MDI. Effect of tyre spoke designs on damage related factors.

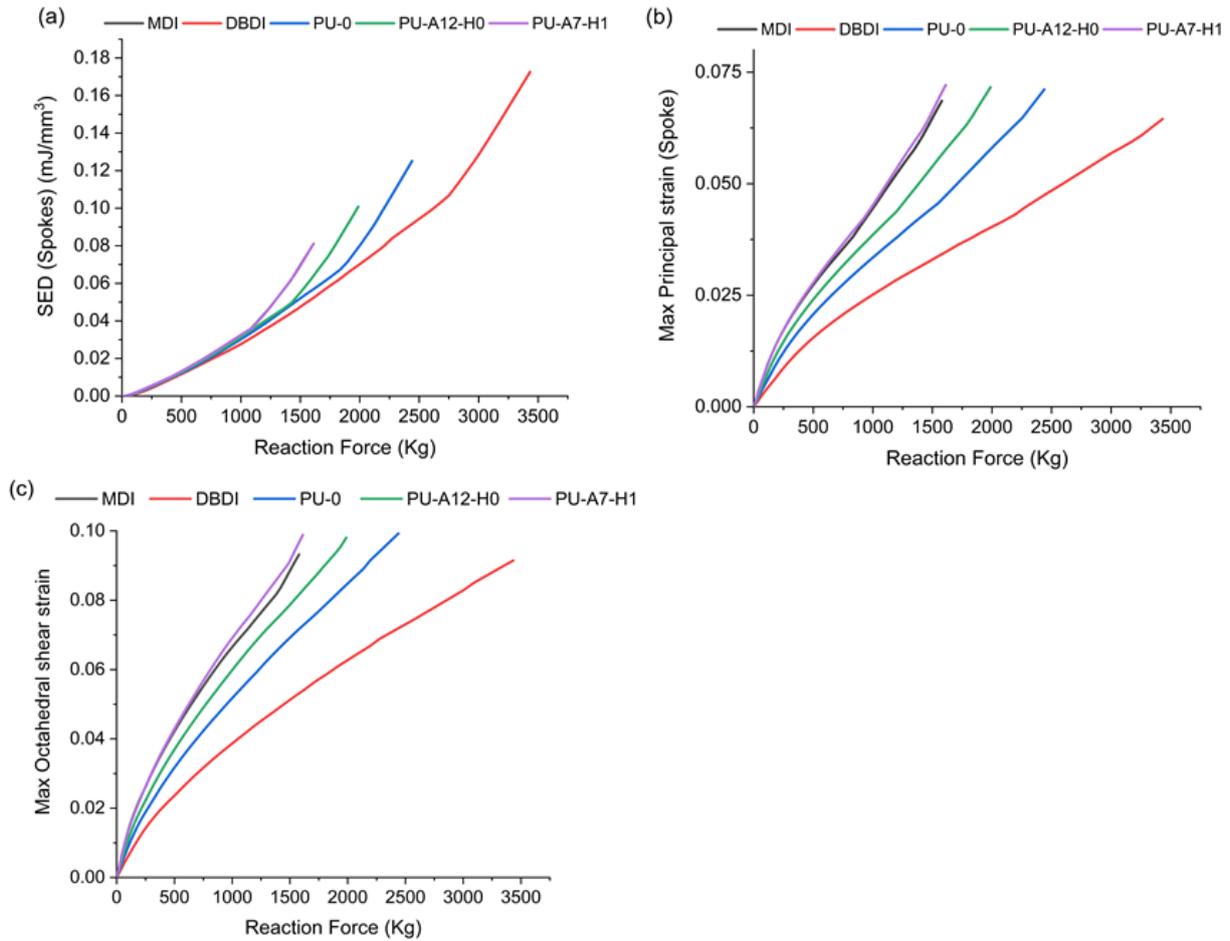

Figure 8 Variation of SED, maximum principal strain and octahedral shear strain with increasing static load for Tweel spokes

In Figs. 9 and 10 the effect of spoke structure on the damage parameters (i.e. SED and maximum principal strain) have been compared. From Fig 9c it can be seen that the value of SED for Honeycomb spoke is almost equal for all five-material property. But there is marginal difference in the maximum principal strain for spokes in NPT-Honeycomb spoke structure. Although there is significant variation in the material property of MDI & PU-A7-H1, The curves of SED and Max principal strain for MDI and PU-A7-H1 are overlapping to each other. Fig. 9 shows that the damage parameter SED alone cannot distinguish between the different material properties for damage behaviour for any particular NPT spoke design. So, in Fig. 10 the other distinguishing damage parameter maximum principal strain has been plotted for all 5 material properties applied for all the tyres. This max principal strain has much variability to capture for damage aspects for all spoke designs than the previously discussed parameter SED. From Fig. 10 it can be seen that the honeycomb spokes have "S-type" variation in the maximum principal strain with increasing load, which is due to their complex design and is quite different than Tweel and UPTIS spokes. If we compare Fig 9 and Fig 10, we can say that combination of Tweel with DBDI has lowest value of SED and maximum principal strain, hence this is the most suitable design for better durability performance. However, from damage perspective, the material property of DBDI appears to be the best material candidate among the 5 materials for all the spoke designs involved.

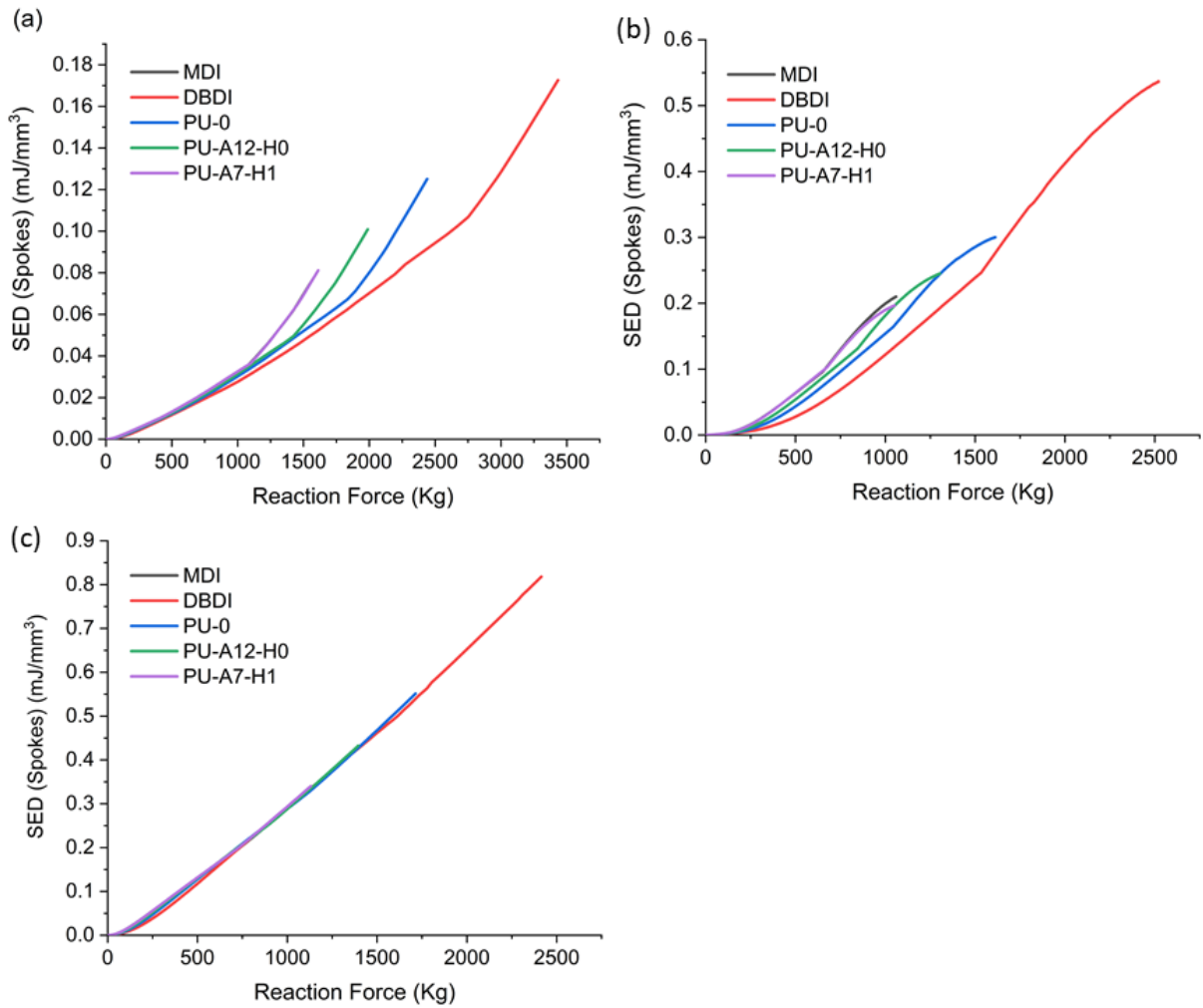

Figure 9 SED variation with increasing static load having variety of PU property for (a) Tweel (b) UPTIS and (c) Honeycomb spokes

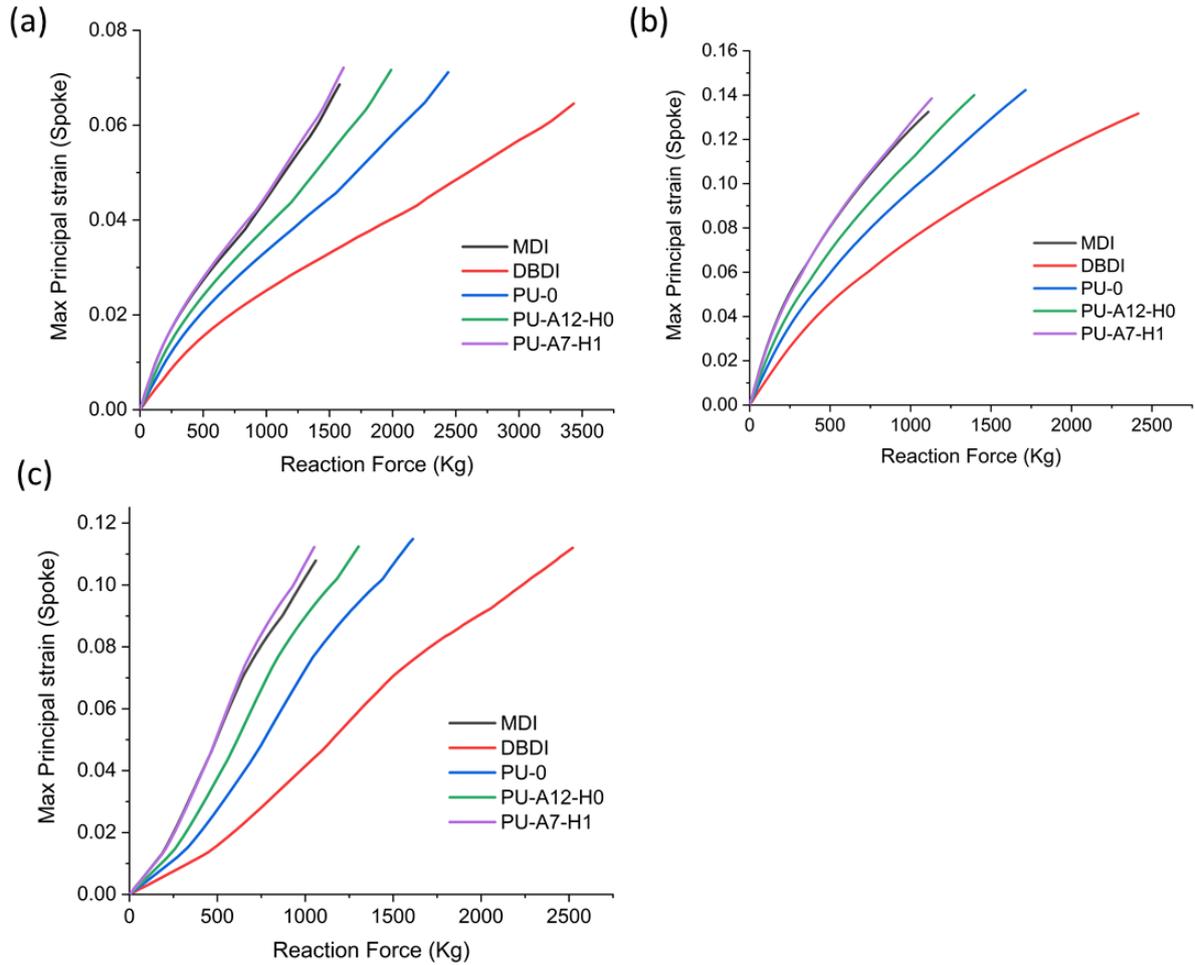

Figure 10 Max principal strain variation with increasing static load having variety of PU property for (a) Tweel (b) UPTIS and (c) Honeycomb spokes

## 4 Conclusion:

FEM models of three different spokes structures (Tweel, Honeycomb and UPTIS) have been developed. Effects of the nonlinearity of PU materials (Linear elastic, MDI, DBDI, PU-0, PU-A12-H0, PU-A7-H1) properties of spokes have been studied with a series of FEM simulations carried out in ANSYS 16.0. The major findings of this study are as follows.

1. Mooney Rivlin 5 parameter model is best to capture all five types of PU hyperelastic behaviour like the non-linearity and rubber modulus changes based on the 5 different nonlinear stress-strain graphs from the materials related to this study.

2. The tyre spoke designs have some degree of effects on the overall tyre stiffness found in the force-displacement curves. But the difference of overall tyre stiffness goes down with increasing load.

3. NPT-UPTIS has softer cushioning effects than the NPT-Honeycomb and NPT-Tweel for both linear and non-linear material properties for spoke structure.

4. Non-linearity of PU-stress strain curve has a larger impact than the 300% rubber modulus (E) on the overall tyre stiffness which is observed for all three spoke designs.

5. Damage predictors like max principal strain and SED are significantly lower in NPT-Tweel than NPT-UPTIS and NPT-Honeycomb. DBDI material has lowest value of SED and Max principal strain in all there spoke structure. Hence it is the best choice for having Tweel with DBDI spoke material, if tire designer wants good tire stiffness and durability performance. However, if the importance is given more on the comfort of driving and more cushioning effect needed, then NPT-UPTIS can be a better choice.

6. Damage predictors like maximum principal strain and octahedral strain have more differentiating effect than strain energy density. SED has very similar effect for all non-linear material property for loading up to 1000 kg. But maximum principal strain has more effect as a damage initiation factor on all nonlinear materials. Based on that it can be concluded that DBDI has highest resistance to crack initiation in the NPT spoke structure.

7. Max-principal strain and max Octahedral shear strains have larger dependence on the material properties. So, these parameters rather may be taken as more important damage parameters for damage related performance prediction.